\documentstyle[epsfig]{aipproc}
\newcommand{\bq}{\begin{equation}}
\newcommand{\eq}{\end{equation}}

\begin{document}
\begin{flushleft}
DESY 97-106 \hfill {\tt hep-ph/9706371} \\
WUE--ITP--97-017 \hfill June 1997 \\
\end{flushleft}

\title{Small-{\boldmath $x$} Resummations for the  \\[1.5mm]
Structure Functions {\boldmath $F_2^{\, p}$}, {\boldmath $F_L^{\, p}$} 
and {\boldmath $F_2^{\,\gamma \,}$}\footnote{
Talk presented by J. Bl\"umlein. To appear in: Proceedings of
the 5th  International Workshop on Deep Inelastic Scattering and QCD 
(DIS$\,$97), Chicago, April 1997}}
 
\author{J. Bl\"umlein$^*$ and A. Vogt$^{\dagger}$}
\address{$^*$DESY--Zeuthen, Platanenallee 6, D--15735 Zeuthen,
 Germany \\
$^{\dagger}$
Institut f\"ur Theoretische Physik, Universit\"at W\"urzburg,
Am Hubland, \\ D--97074  W\"urzburg, Germany}

\maketitle

\begin{abstract}
The numerical effects of the known all-order leading and next-to-leading
logarithmic small-$x$ contributions to the anomalous dimensions and 
coefficient functions of the unpolarized singlet evolution are discussed
for the structure functions $F_2^{\, ep}(x,Q^2)$, $F_L^{\, ep}(x,Q^2)$, 
and $F_2^{\, e\gamma}(x,Q^2)$.
\end{abstract}
 
%%%%%%%%%%%%%%%%%%%%%%%%%%%%%%%%%%%%%%%%%%%%%%%%%%%%%%%%%%%%%%%%%%%%%%%%
\vspace*{-4mm}
\subsection*{Introduction}

\noindent
The evolution kernels of the deep-inelastic scattering (DIS) structure
functions contain large logarithmic contributions for small Bjorken-$x$.
The effect of resumming these terms to all orders in $\alpha_s$ can be 
consistently studied in a framework based on the renormalization group 
(RG) equations, which describes the mass factorization. In this 
framework, the evolution equations of fixed-order perturbative QCD are 
generalized by including the resummed small-$x$ contributions to the
respective anomalous dimensions and Wilson coefficients~\cite
{BV:LI,BV:CH,BV:CC,BV:NSP} beyond next-to-leading order in $\alpha_s$ 
(NLO). The numerical impact of these higher-order contributions has 
been investigated for the non-singlet nucleon structure functions 
$F_2^{\,p-n} $ and $F_3^{\,\nu N}$~\cite{BV:NS}, $g_1^{\, p-n}$~\cite
{BV:NS,BV:KOD} and $g_5^{\,\gamma Z}$~\cite{BV:BV1}; for the polarized 
singlet quantity $g_{1,\rm S}^{\, p}$~\cite{BV:POL}, and for the 
unpolarized singlet structure functions $F_{2,\rm S}$~\cite
{BV:F2,BV:BV2,BV:BO} and $F_{L,\rm S}^{\, p}$ \cite{BV:BV2}.
$F_{2,\rm S}^{\, p}$ and $F_{L,\rm S}^{\, p}$ have been studied using 
different RG-based approaches as well \cite{BV:TH}.

In the present note we extend a previous account~\cite{BV:BV1} by 
considering, besides the resummed next-to-leading logarithmic small-$x$ 
(NL$x$) quark terms of ref.~\cite{BV:CH}, also the recently derived 
NL$x$ contributions $\propto N_f$ to the anomalous dimension 
$\gamma_{gg}$~\cite{BV:CC} and their impact on $F_2^{\, p}$. 
Furthermore, we briefly discuss the numerical resummation effects on 
the evolution of $F_L^{\, p}$ and the photon structure function 
$F_2^{\, \gamma}$. Details of the calculations may be found in 
ref.~\cite{BV:BV2}.

%%%%%%%%%%%%%%%%%%%%%%%%%%%%%%%%%%%%%%%%%%%%%%%%%%%%%%%%%%%%%%%%%%%%%%%%
\subsection*{The NL{\boldmath $x$} Contributions {\boldmath $\propto 
N_f$} to {\boldmath $\gamma_{gg}$}}

\noindent
These terms were calculated in ref.~\cite{BV:CC}. In the $\overline
{\rm MS}$--DIS scheme they read~\cite{BV:BV2}
%-----------------------------------------------------------------------
\begin{eqnarray}
\label{BV:E:cc}
 \gamma_{gg,\rm NL}^{q\overline{q}, {\rm DIS}} &=&
 \gamma_{gg,\rm NL}^{q\overline{q},{Q_0}}\, + \,\frac{\beta_0}{4\pi }\, 
 \alpha_s^2 \frac{d \ln R(\alpha_s)}{d \alpha_s}\, + \,\frac{C_F}{C_A}
 \left[1 - R(\alpha_s) \right] \gamma_{qg,\rm NL}^{ {Q_0}} 
 \nonumber\\
&\equiv &
 \alpha_s \sum_{k=1}^{\infty} \left[ \frac{N_f}{6\pi} \left( 
 d_{gg,k} ^{\, q\overline{q},\, (1)} + \frac{C_F}{C_A} d_{gg,k}^{\, 
 q\overline{q},\, (2)} \right) + \frac{\beta_0}{4 \pi} \hat{r}_k \right]
 \left( \frac{\overline{\alpha}_s}{N\! -\! 1} \right)^{k-1}~,
\end{eqnarray}
%-----------------------------------------------------------------------
with $\gamma_{gg,\rm NL}^{q\overline{q},Q_0}$ being the $N_f$ 
contribution in the $Q_0$ scheme~\cite{BV:MC}. 
$N$ denotes the usual Mellin variable,
$\overline{\alpha}_s \equiv C_A \,\alpha_s / \pi$, and $R(\alpha_s)$ 
is defined in ref.~\cite{BV:CH}. $\gamma_{gg,\rm NL}^{q\overline{q}}$ 
contains terms $\propto C_F/C_A$ in both schemes, whereas the $\beta_0
$-contribution originates in transformation from the $Q_0$ scheme
to the $\overline{\rm MS}$--DIS scheme.
Numerical values for the coefficients $d_{gg,k}^{\,q\overline{q},\,
(1,2)} $ and $\hat{r}_k$ are given in Table~1.

%-----------------------------------------------------------------------
\vspace*{4mm}
\begin{center}
\small
\begin{tabular}{||r||r|r|r||}
\hline\hline
 & & & \\[-4mm]
\multicolumn{1}{||c||}{$k$} &
\multicolumn{1}{c|}{$d_{gg,k}^{\,q\overline{q},\, (1)}$} &
\multicolumn{1}{c|}{$d_{gg,k}^{\,q\overline{q},\, (2)}$} &
\multicolumn{1}{c||}{$\hat{r}_k$} \\
 & & & \\[-4mm] \hline\hline
 & & & \\[-4mm] 
%-----------------------------------------------------------------------
1 & --1.000000000$\,$E+0 & 0.000000000$\,$E+0 &   0.000000000$\,$E+0 \\
2 & --3.833333333$\,$E+0 & 0.000000000$\,$E+0 &   0.000000000$\,$E+0 \\
3 & --2.299510376$\,$E+0 & 0.000000000$\,$E+0 &   0.000000000$\,$E+0 \\
4 & --5.065605818$\,$E+0 & 3.205485075$\,$E+0 &   9.616455224$\,$E+0 \\
5 & --3.523670351$\,$E+1 & 8.568702514$\,$E+0 & --3.246969702$\,$E+0 \\
6 & --3.218245315$\,$E+1 & 1.835447655$\,$E+1 &   2.281241061$\,$E+1 \\
7 & --1.060268680$\,$E+2 & 8.632838009$\,$E+1 &   1.654162989$\,$E+2 \\
8 & --4.853159484$\,$E+2 & 1.924088636$\,$E+2 & --2.469139930$\,$E+0 \\
9 & --5.806186371$\,$E+2 & 4.962344972$\,$E+2 &   7.458249428$\,$E+2 \\
10& --2.176371931$\,$E+3 & 1.794742819$\,$E+3 &   2.784859262$\,$E+3 \\
%-----------------------------------------------------------------------
11& --7.553679737$\,$E+3 & 4.023320193$\,$E+3 &   1.505001272$\,$E+3 \\
12& --1.158215080$\,$E+4 & 1.136559381$\,$E+4 &   1.818320928$\,$E+4 \\
13& --4.328579102$\,$E+4 & 3.589638820$\,$E+4 &   4.899274185$\,$E+5 \\
14& --1.269309428$\,$E+5 & 8.412529889$\,$E+4 &   6.109247725$\,$E+5 \\
15& --2.392549581$\,$E+5 & 2.456097133$\,$E+5 &   3.984470167$\,$E+5 \\
16& --8.469557573$\,$E+5 & 7.168572021$\,$E+6 &   9.205515787$\,$E+5 \\
17& --2.262541206$\,$E+6 & 1.764587230$\,$E+6 &   1.783326920$\,$E+6 \\
18& --4.974873276$\,$E+6 & 5.167844173$\,$E+6 &   8.347774614$\,$E+6 \\
19& --1.648990863$\,$E+7 & 1.443009883$\,$E+7 &   1.842662795$\,$E+7 \\
20& --4.222994214$\,$E+7 & 3.702246358$\,$E+7 &   4.535538189$\,$E+7 \\
%-----------------------------------------------------------------------
[1mm] \hline \hline
\end{tabular}
\vspace{4mm}
\normalsize
{\sf Table~1:~~Numerical values of the expansion coefficients for
 $\gamma_{gg,\rm NL}^{q\overline{q},{\rm DIS}}$ in eq.~(\ref{BV:E:cc}).}
\end{center}
%-----------------------------------------------------------------------

%%%%%%%%%%%%%%%%%%%%%%%%%%%%%%%%%%%%%%%%%%%%%%%%%%%%%%%%%%%%%%%%%%%%%%%%
\subsection*{Less Singular Small-{\boldmath $x$} Contributions to 
 {\boldmath ${\gamma}$}}

\noindent
The small-$x$ resummed anomalous dimension matrix $\hat{\gamma}^{\,\rm 
res}$ does not comply with the energy-momentum sum rule for the parton 
densities. Several prescriptions have been imposed for restoring this 
sum rule beyond NLO \cite{BV:F2,BV:BV2,BV:BO}, e.g.,
%-----------------------------------------------------------------------
\begin{equation}
\begin{array}{cl}
{\rm A:} & \hat{\gamma}^{\,\rm res} (n, \alpha_s) \rightarrow  
 \hat{\gamma}^{\,\rm res}(n, \alpha_s) - 
 \hat{\gamma}^{\,\rm res}(0, \alpha_s) \\
{\rm B:} & \hat{\gamma}^{\,\rm res} (n, \alpha_s) \rightarrow  
 \hat{\gamma}^{\,\rm res}(n, \alpha_s)\, (1 - n) \\
{\rm D:} & \hat{\gamma}^{\,\rm res} (n, \alpha_s) \rightarrow 
 \hat{\gamma}^{\,\rm res}(n, \alpha_s)\, (1 - 2n + n^3) \:\: .
\end{array}
\end{equation}
%-----------------------------------------------------------------------
The difference between the results obtained with these prescriptions
allows for a rough estimate of the possible effect of the presently
unknown higher-order terms less singular at small-$x$ ($ n \equiv N\! 
-\! 1 \rightarrow 0$).

%%%%%%%%%%%%%%%%%%%%%%%%%%%%%%%%%%%%%%%%%%%%%%%%%%%%%%%%%%%%%%%%%%%%%%%%
\subsection*{The Resummed Evolution of {\boldmath $F_2^{\, ep}$} and 
 {\boldmath $F_L^{\, ep}$}}

\noindent
The numerical effect of the known small-$x$ resummations on the 
behavior of the proton structure functions $F_2$ and $F_L$ is 
illustrated in Fig.~1. For both the NLO and the resummed calculations, 
the MRS(A$^{\prime}$) DIS-scheme parton densities have been employed as 
initial distributions at $Q_0^2 = 4 \mbox{ GeV}^2$, together with
$\Lambda_{\overline{\rm MS}}^{(4)} = 231 \mbox{ MeV}$ \cite{BV:MRS}. 
They behave like $ xg,xq \sim x^{-0.17}$ at small $x$, with the quark 
part rather directly constrained by present HERA $F_2$ data.

%-----------------------------------------------------------------------
\begin{figure}[hb]
\vspace{-7mm}
\begin{center}
\mbox{\hspace*{-2mm}\epsfig{file=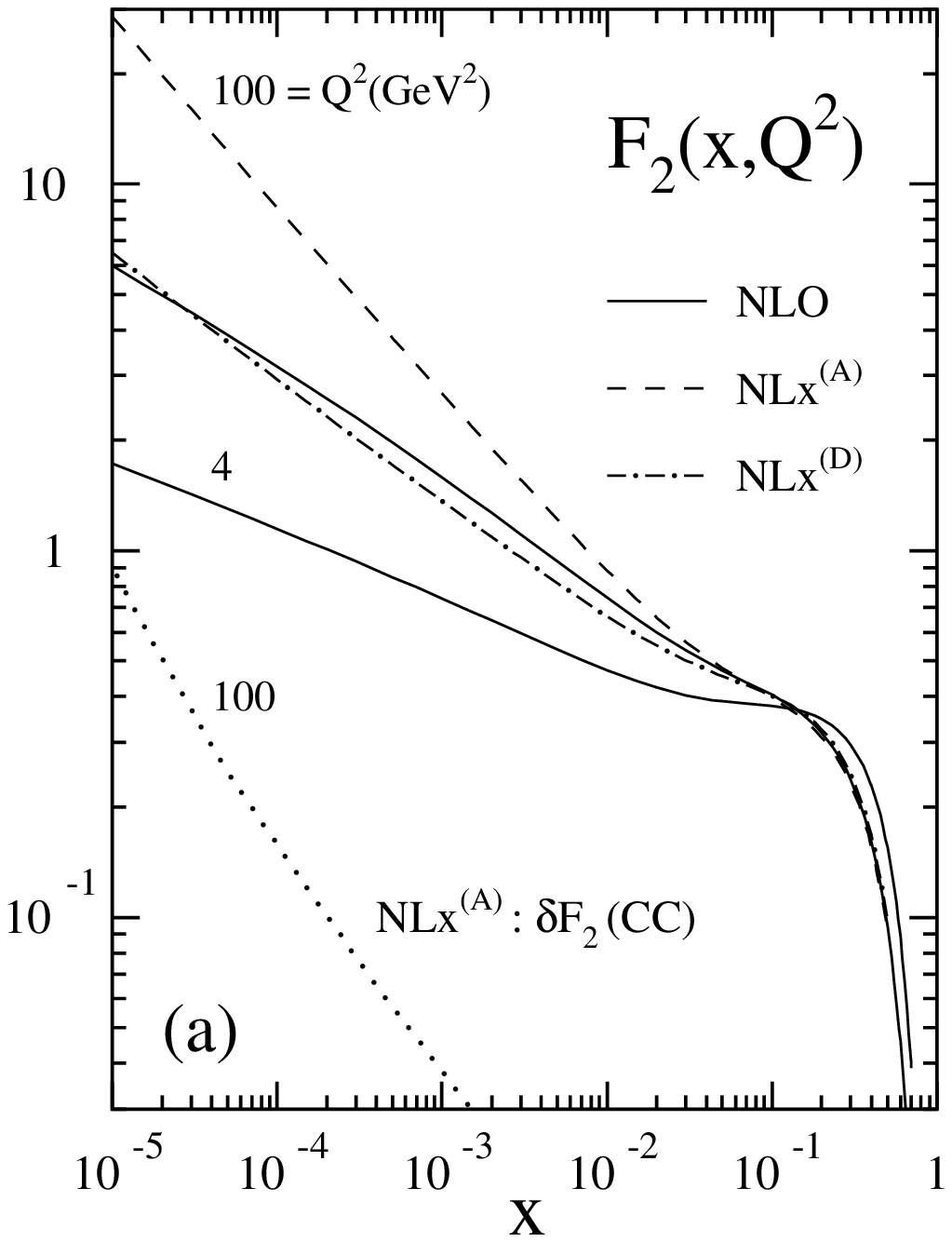,width=7cm}\hspace*{-2mm}
      \epsfig{file=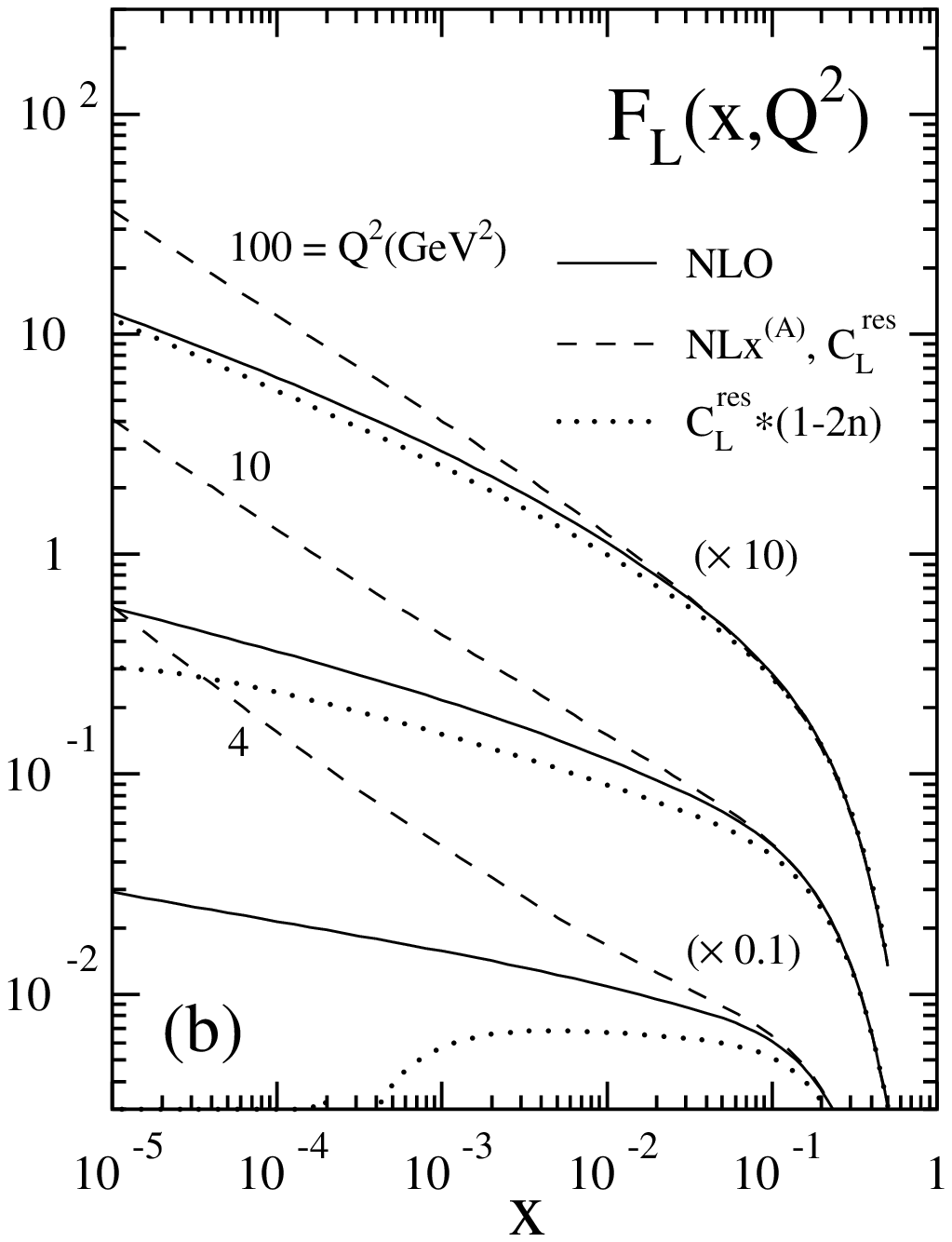,width=7cm}}
\vspace{-5mm}
\end{center}
{\sf Figure~1: The resummed small-$x$ evolutions of the proton structure
 functions $F_2$ and $F_L$ compared to the NLO results. The dotted curve
 in the $F_2$ part represents the contribution of $\gamma_{gg,\,\rm NL}
 ^{q\overline{q},\, {\rm DIS}}$ only. The possible impact of (presently 
 unknown) less singular higher-order terms is indicated, cf.~eq.~(2) and
 the discussion in the text.}
\vspace*{-7mm}
\end{figure}
%-----------------------------------------------------------------------

The resummation effects on $F_2(x,Q^2)$ at small $x$ are displayed in 
Fig.\ 1$\,$(a). Note the huge effect arising from the NL$x$ quark 
anomalous dimensions \cite{BV:CH} and its large uncertainty due to 
unknown less singular terms. The impact of $\gamma_{gg,\rm NL}
^{q\overline{q}}$~\cite {BV:CC} is displayed separately. It amounts to
less than 3\% over the full $x$-range shown. It will be interesting to 
see to which extent the forthcoming complete NL$x$ anomalous dimensions 
\cite{BV:MCI} will modify these results.

\vspace{1mm}
The longitudinal structure function $F_L(x,Q^2)$ is considered in 
Fig.\ 1$\,$(b). Obviously substantial contributions can also be 
expected from subleading small-$x$ terms in the coefficient functions 
$C_L$. In fact, these uncertainties are large.  Thus both for the 
small-$x$ resummed contributions to anomalous dimensions and coefficient
functions further subleading terms need to be calculated. Further 
insight into the interplay of leading and less singular terms in $N$ 
may also be gained from the structure of the fixed-order anomalous 
dimensions and coefficient functions. Besides the known NLO result, 
particularly the yet unknown 3--loop anomalous dimensions are of 
interest here.

%%%%%%%%%%%%%%%%%%%%%%%%%%%%%%%%%%%%%%%%%%%%%%%%%%%%%%%%%%%%%%%%%%%%%%%%
\subsection*{The Resummation of the Small-{\boldmath $x$} Contributions 
 to {\boldmath $F_2^{\, \gamma}$}}

\noindent
The evolution of the photon structure functions is, at the lowest order
in $\alpha_{\rm em}$ considered here, governed by an inhomogeneous 
generalization of the hadronic evolution equations. At the present 
resummation accuracy \cite{BV:LI,BV:CH} the additional anomalous 
dimensions $\gamma_{q \gamma}$ and $\gamma_{g \gamma}$ do not receive 
any non-vanishing higher-order small-$x$ contributions \cite{BV:BV2}. 
Hence the resummation effect on the photon-specific inhomogeneous 
solution originates solely from the resummed homogeneous evolution 
operator.
%-----------------------------------------------------------------------
\begin{figure}[bht]
\vspace{-7mm}
\begin{center}
\mbox{\hspace*{-2mm}\epsfig{file=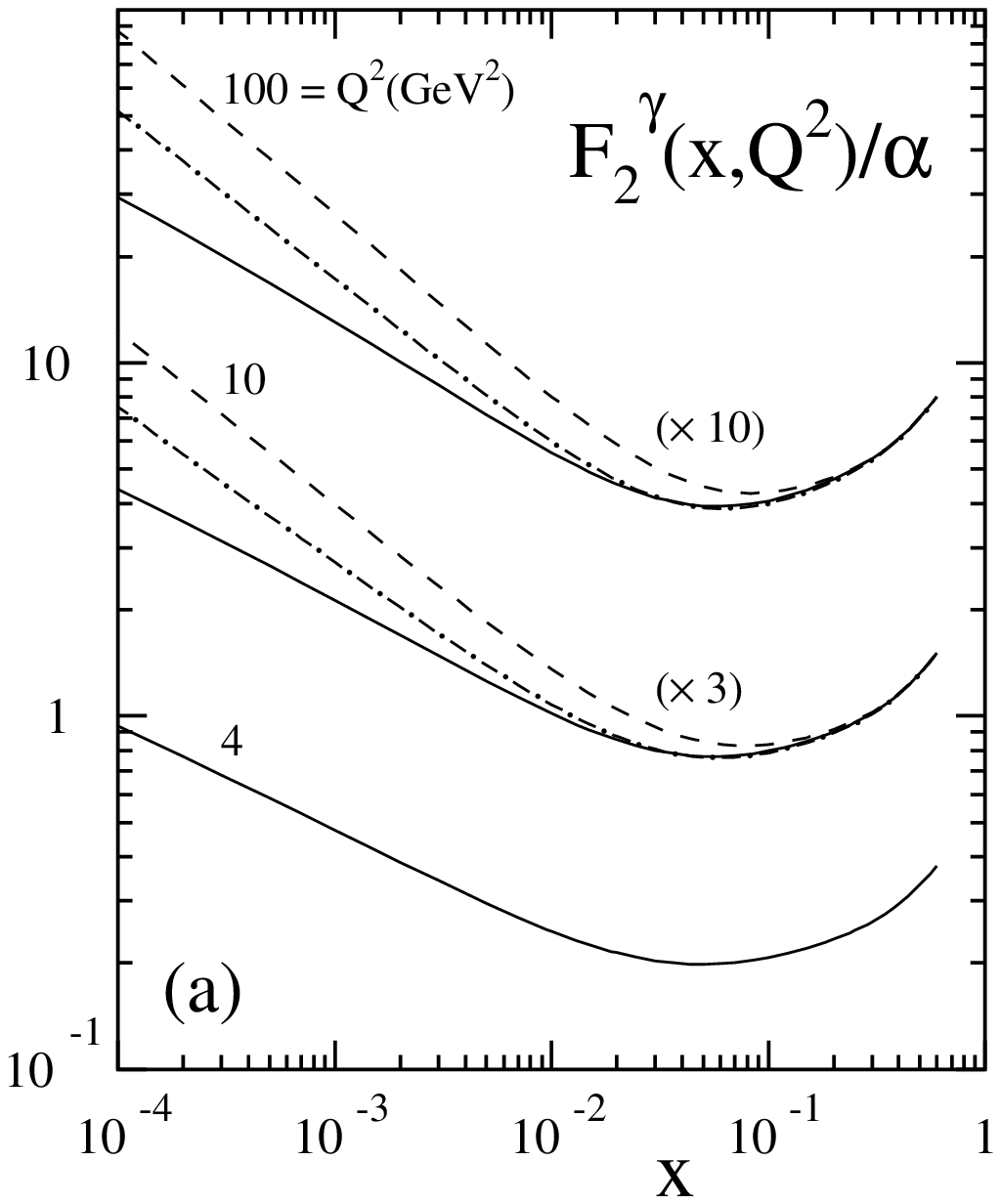,width=7cm}\hspace*{-2mm}
      \epsfig{file=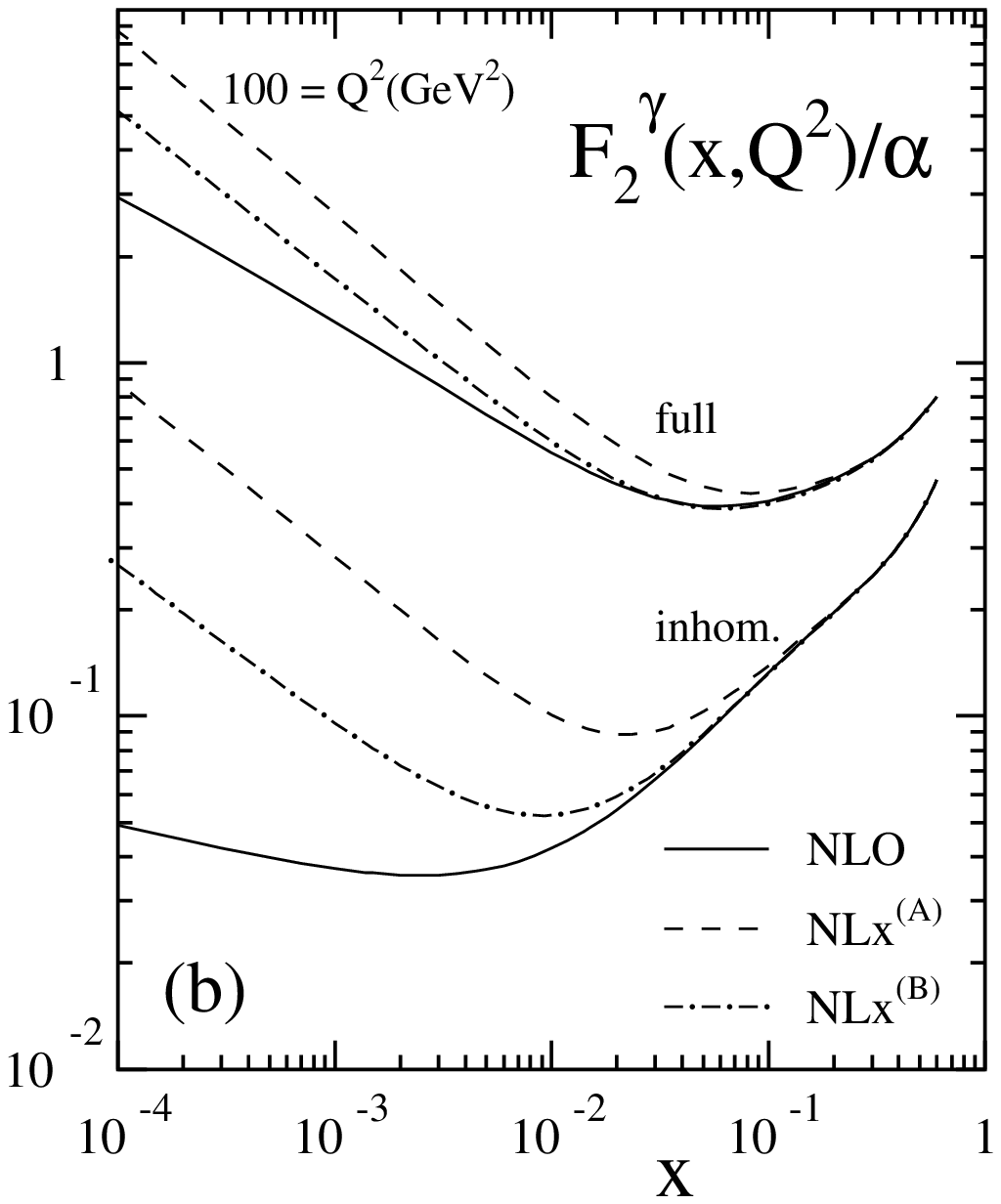,width=7cm}}
\vspace{-4mm}
\end{center}
{\sf Figure~2: The small-$x$ evolution of the photon structure function 
 $F_2^{\,\gamma }$ in NLO and using the NL$x$ resummed anomalous
 dimensions.}
\vspace*{-7mm}
\end{figure}
%-----------------------------------------------------------------------

The resummed evolution of the structure function $F_2^{\,\gamma}$ is
compared to the NLO results in Fig.\ 2. The NLO GRV parametrization 
has been used for the initial distributions at $Q_0^2 = 4\mbox{ GeV}^2$,
together with $\Lambda_{\overline{\rm MS}}^{(4)} = 200 \mbox{ MeV}$ 
\cite{BV:GRV}. The overall small-$x$ behavior, presented in Fig.\ 2$\,
$(a), is rather similar to the hadronic case, due to the dominance of
the homogeneous solution. Note, however, the significantly enhanced
resummation effect in the inhomogeneous solution separately shown in 
Fig.\ 2$\,$(b). This behavior is dominated by the convolution of the 
resummed hadronic evolution operator with the leading-order 
photon-quark anomalous dimension, which, unlike the hadronic initial
distributions, is large for $x \rightarrow 1$.

\vspace{5mm}
\noindent
{\bf Acknowledgement:} This work was supported in part by the German
Federal Ministry for Research and Technology (BMBF) under contract 
No.\ 05~7WZ91P~(0).


\begin{references}
%
%[1]
\bibitem{BV:LI}
Y. Balitsky and L. Lipatov, {\it Sov.\ J.\ Nucl.\ Phys.\ }{\bf 28} 822 
 (1978).
%
%[2]
\bibitem{BV:CH}
S. Catani and F. Hautmann, {\it Nucl.\ Phys.\ }{\bf B427} 475 (1994).
%
%[3]
\bibitem{BV:CC}
G. Camici and M. Ciafaloni, {\it Phys.\  Lett.\ }{\bf B386} 341 (1996); 
 {\tt hep-ph/9701303}.
%
%[4]
\bibitem{BV:NSP}
R. Kirschner and L. Lipatov, {\it Nucl.\ Phys.\ }{\bf B213} 122 
 (1983);\\
J. Bartels, B. Ermolaev, and M. Ryskin, {\it Z. Phys.\ }{\bf C72} 627 
 (1997).
%
%[5]
\bibitem{BV:NS}
J. Bl\"umlein and A. Vogt, {\it Phys.\ Lett.\ }{\bf B370} 149 (1996);\\
{\it Acta Phys.\ Polonica} {\bf B27} 1309 (1996).
%
%[6]
\bibitem{BV:KOD}
J. Kiyo, J. Kodaira, and H. Tochimura, {\tt hep-ph/9701365}.
%
%[7]
\bibitem{BV:BV1}
J. Bl\"umlein, S. Riemersma, and A. Vogt, 
 {\it Nucl. Phys. {\bf B} (Proc.\ Suppl.)} {\bf 51C} 30 (1996).
%
%[8]
\bibitem{BV:POL}
J. Bl\"umlein and A. Vogt, {\it Phys.\ Lett.\ }{\bf B386} 350 (1996).
%
%[9]
\bibitem{BV:F2}
R.K. Ellis, F. Hautmann, and B. Webber, {\it Phys.\ Lett.\ }{\bf B348} 
 582 (1995);\\
R. Ball and S. Forte, {\it Phys.\ Lett.\ }{\bf B351} 313 (1995), 
 {\bf B358} 365 (1995).
%
%[10]
\bibitem{BV:BV2}
J. Bl\"umlein and A. Vogt, DESY 96--096.
%
%[11]
\bibitem{BV:BO}
I. Bojak and M. Ernst, {\it Phys. Lett.\ }{\bf B397} 296 (1997); 
 {\tt hep-ph/9702282}.
%
%[12]
\bibitem{BV:TH}
J. Forshaw, R. Roberts, and R. Thorne, {\it Phys.\ Lett.\ }{\bf B356} 
 79 (1995);\\
R. Thorne, {\it Phys. Lett.\ }{\bf B392} 463 (1997); 
 {\tt hep-ph/9701241}.
%
%[13]
\bibitem{BV:MC}
M. Ciafaloni, {\it Phys.\ Lett.\ }{\bf B356} 74 (1995).
%
%[14]
\bibitem{BV:MRS}
A.D. Martin, R.G. Roberts, and W.J. Stirling, 
 {\it Phys.\ Lett.\ }{\bf B354} 155 (1995)
%
%[15]
\bibitem{BV:MCI}
M. Ciafaloni et al., in preparation.
%
%[16]
\bibitem{BV:GRV}
M. Gl\"uck, E. Reya, and A. Vogt, {\it Phys.\ Rev.\ }{\bf D46} 1973 
 (1992).
%
\end{references}
\end{document}